\documentclass[aps,prb,twocolumn,superscriptaddress,groupedaddress,showpacs,floatfix,amsmath,amssymb]{revtex4}
\usepackage{graphicx,xcolor}
\def\etal{{\em{et al.}}}

\begin{document}

\title{Study of off-diagonal disorder using the typical medium dynamical cluster approximation }

\author{H.\ Terletska}
\email{terletska.hannas@gmail.com}
\affiliation{Department of Physics \& Astronomy, Louisiana State University, Baton Rouge, Louisiana 70803, USA}
\affiliation{Center for Computation \& Technology, Louisiana State University, Baton Rouge, Louisiana 70803, USA}

\author{C. E.\ Ekuma}
\affiliation{Department of Physics \& Astronomy, Louisiana State University, Baton Rouge, Louisiana 70803, USA}
\affiliation{Center for Computation \& Technology, Louisiana State University, Baton Rouge, Louisiana 70803, USA}

\author{C.\ Moore}
\affiliation{Department of Physics \& Astronomy, Louisiana State University, Baton Rouge, Louisiana 70803, USA}
\affiliation{Center for Computation \& Technology, Louisiana State University, Baton Rouge, Louisiana 70803, USA}

\author{K.-M.\ Tam}
\affiliation{Department of Physics \& Astronomy, Louisiana State University, Baton Rouge, Louisiana 70803, USA}
\affiliation{Center for Computation \& Technology, Louisiana State University, Baton Rouge, Louisiana 70803, USA}

\author{J.\ Moreno}
\affiliation{Department of Physics \& Astronomy, Louisiana State University, Baton Rouge, Louisiana 70803, USA}
\affiliation{Center for Computation \& Technology, Louisiana State University, Baton Rouge, Louisiana 70803, USA}

\author{M.\ Jarrell}
\email{jarrellphysics@gmail.com}
\affiliation{Department of Physics \& Astronomy, Louisiana State University, Baton Rouge, Louisiana 70803, USA}
\affiliation{Center for Computation \& Technology, Louisiana State University, Baton Rouge, Louisiana 70803, USA}

\date{\today}

\begin{abstract}
We generalize the typical medium dynamical cluster approximation (TMDCA) and the local Blackman, Esterling, and 
Berk (BEB) method for systems with off-diagonal disorder.  Using our extended formalism
we perform a systematic study of the effects of non-local disorder-induced correlations and of 
off-diagonal disorder on the density of states and the mobility edge of the Anderson localized states.
We apply our method to the three-dimensional Anderson model with configuration dependent hopping and find fast 
convergence with modest cluster sizes. Our results are in good agreement with the data obtained using exact 
diagonalization, and the transfer matrix and kernel polynomial methods.

\end{abstract}

\pacs{71.27.+a, 02.70.-c, 71.10.Fd, 71.23.An}
%  Computational techniques; simulations: 02.70.-c
%  Strongly correlated electron systems, 71.27.+a
%  Insulator-metal transitions, 71.30.+h
%  Anderson localization: disordered solids, 71.23.An

\maketitle

\section{Introduction}
\label{sec:intro}
Disorder which is inevitably present in most materials can dramatically
affect their properties.~\cite{Lee_RevModPhys,Belitz_RevModPhys} It can lead to 
changes in their electronic structure and transport. One of the most interesting effects 
of disorder is the spatial confinement of charge carriers due to coherent backscattering
off random impurities which is known as Anderson localization.~\cite{Anderson,PhysRevLett.42.673}
Despite progress over the last decades, the subject of Anderson localization 
remains an active area of research. The lack of quantitative analytical results
has meant that numerical investigations (see for e.g., 
Refs.~\onlinecite{Bulka85,Bulka87,Slevin2014,PhysRevLett.78.4083,Slevin99,PhysRevLett.105.046403,PhysRevB.84.134209}) 
have provided a significant role in 
understanding the Anderson transition.~\cite{Nakayama-Yakubo-2003,Markos-review-2006,0305-4470-33-42-103}

The simplest model used to study the effects of disorder in materials is 
a single band tight binding model with a random on-site disorder potential.~\cite{a_gonis_92}
Such a model is justified when the disorder is introduced by substitutional impurities,
as in a binary alloy. The substitution of host atoms by impurities 
only leads to changes of the local potential on the substitutional site and, on 
average, does not affect the neighbors.~\cite{a_gonis_92,soven_67} In this situation, 
the disorder appears only in the diagonal terms of the Hamiltonian and hence is 
referred to as diagonal disorder. However, when the bandwidth 
of the dopant is very different from the one
of the pure host, such substitution results not only in the change of the local
potential but may also affect the neighboring sites.~\cite{a_gonis_92}
Consequently, a simple model to capture such effects should include both  
random local potentials and random hopping amplitudes which depend on the 
occupancy of the sites.  The dependence of the hopping amplitude
on the disorder configuration is usually referred to as off-diagonal disorder.
Of course, a proper theoretical description of realistic disordered materials 
requires the inclusion of both diagonal and off-diagonal randomness. 

The coherent potential approximation (CPA) is a widely used single site mean 
field theory for systems with strictly diagonal disorder.~\cite{soven_67}
Blackman, Esterling and Berk (BEB)~\cite{PhysRevB.4.2412} have extended the 
CPA to systems with off-diagonal disorder. However, being single-site approximations, 
the CPA and the BEB theories neglect all disorder induced non-local correlations.

There have been a number of attempts to develop systematic nonlocal extensions 
to the CPA. These include cluster extensions such as the molecular coherent potential 
approximation (MCPA),~\cite{gonis_mcpa,gonis_odd1} the dynamical cluster 
approximation (DCA),~\cite{Hettler98,Hettler00,m_jarrell_01a} etc. 
Self-consistent mean field studies of off-diagonal disorder have been conducted
by a number of authors.~\cite{Shiba, Brezini-1, Brezini-2, gonis_odd1}
However, all these studies have been performed at the local single-site BEB level.
To include the effects of off-diagonal disorder, Gonis \cite{gonis_mcpa} extended the Molecular CPA, 
which uses a self-consistently
embedded finite size cluster to capture non-local corrections to the CPA.
However, he criticized the MCPA 
for violating  translational invariance and other critical properties
of a valid quantum cluster theory.~\cite{a_gonis_92,PhysRevB.89.081107}  In order to take into account such 
non-local effects on off-diagonal disorder models while maintaining translational invariance, we extend the 
BEB formalism using the DCA scheme.~\cite{Hettler98,Hettler00,m_jarrell_01a} 

While the CPA, DCA, and BEB have shown to be successful self-consistent mean-field theories
for the quantitative description of the density of states and electronic structure 
of disordered systems, they can not properly address the physics of  Anderson 
localization. These mean field approaches describe the effective medium using the 
average density of states which is not critical at the 
transition.~\cite{Thouless,PhysRevB.89.081107,Nakayama-Yakubo-2003,Abou-Chacra}
Thus, theories which rely on such averaged quantities will fail to properly 
characterize Anderson localization. As noted by Anderson, the probability distribution 
of the local density of states must be considered, focusing on 
the most probable or the \emph{typical} value.~\cite{Anderson,anderson1978local}
Close to the Anderson transition, the distribution is found to have very long tails 
characteristic of a log-normal distribution.\cite{PhysRevLett.105.046403,PhysRevLett.72.526,Vollhardt}  
In fact, the distribution is log-normal up to ten orders of magnitude~\cite{PhysRevB.81.155106} 
and so the typical value~\cite{Janssen98,Crow1988,Vollhardt,Derrida198429} is the geometrical 
mean. Based on this idea, Dobrosavljevi\'{c} \etal~\cite{Vlad2003} formulated a single site 
typical medium theory (TMT) for the Anderson localization. This approximation gives a qualitative 
description of the Anderson localization in three dimensions. However, it fails 
to properly describe the trajectory of the mobility edge (which separates the extended and localized states) 
as it neglects non-local corrections
and so does not include the effects of coherent backscattering.~\cite{2005PhyB..359..789A} It also 
underestimates considerably the critical strength of the disorder at which the localization happens. 
In addition, TMT is only formulated for diagonal disorder.  

Recently, by employing the DCA within the typical medium analysis, we developed a systematic 
Typical Medium Dynamical Cluster Approximation (TMDCA) formalism.~\cite{PhysRevB.89.081107}
The TMDCA provides an accurate description of the Anderson localization transition for modest
cluster sizes in three-dimensional models with diagonal disorder while recovering the TMT for 
a one-site cluster. In this work, we generalize our recently proposed TMDCA scheme to address 
the question of electron localization in systems with both diagonal and off-diagonal disorder.

To go beyond the local single-site CPA-like level of the BEB formalism, we employ
the DCA~\cite{Hettler98,Hettler00,m_jarrell_01a} scheme which systematically 
incorporates non-local spatial correlations effects. Hence, in this paper, we first present an 
extension of the DCA for systems with both diagonal and off-diagonal disorder. Next, we 
develop a typical medium dynamical cluster approximation formalism capable of incorporating 
the effects of Anderson localization  both for diagonal and off-diagonal disorder. 
We then perform a systematic study of the effects of non-local correlations and off-diagonal 
randomness on the density of states and electron localization. The results of our calculations 
are compared with the ones obtained with other numerical methods for finite size lattices, including
exact diagonalization, kernel polynomial, and transfer matrix methods.

The paper is organized as follows: following the Introduction in Sec.~\ref{sec:intro} 
we present the model and describe the details of the formalism we 
used in Sec.~\ref{sec:formalism}.  In Sec.~\ref{sec:Results_DCA} we present our results of the average density of states 
for both diagonal and off-diagonal disorder cases. In Sec.~\ref{sec:Results_TMDCA}
we consider the effects of diagonal and off-diagonal disorder on the typical 
density of states, from which we extract the mobility edges and construct a complete phase diagram in the disorder-energy
parameter space. We summarize and discuss future directions in Sec.~\ref{sec:conclusion}.

%%%%%%%%%%%%%%%%%%%%%%%%%%%%%%%%%%%%%%%%%%%%%%%%%%%%%%%%%%%%%%%%%%%%%%%%%%%%%%%%%%%%%%%

\section{Formalism}
\label{sec:formalism}

\subsection{Dynamical cluster approximation for off-diagonal disorder}
\label{sec:ODD}
The simplest model widely used to study disordered systems is the
single band tight binding Hamiltonian

\begin{equation}
H=-\sum_{<i,j>}t_{ij}(c_{i}^{\dagger}c_{j}+h.c.)+\sum_{i}v_{i}n_{i},
\label{eq:1}
\end{equation}
where disorder is modeled by a local potential $v_{i}$ which is 
a random variable with probability distribution function $P(v_{i})$.
We will focus on the binary disorder case, where some host $A$ atoms are substituted
with $B$ impurities with a probability distribution function of the form
\begin{equation}
P(v_{i})=c_{A}\delta(v_{i}-V_{A})+c_{B}\delta(v_{i}-V_{B}),
\label{eq:pdf}
\end{equation}
where $c_B=1-c_A$.
For the diagonal disorder case when the bandwidth
of the pure host $A$ is about the same that the bandwidth of the $B$ system, such substitution
results only in a change of the local potential $v_{i}$ at the replaced site
$i$. This corresponds to changes in the diagonal elements of the Hamiltonian. 
In this case it is assumed that substitution of impurity atoms on average has no 
effect on hopping amplitudes to the neighboring atoms.

For systems with off-diagonal disorder, the randomness is introduced
not only locally in the random diagonal potential $v_{i}$, but also through the hopping amplitudes.
To model this, BEB~\cite{PhysRevB.4.2412} introduced the disorder configuration dependent 
hopping amplitude of electrons $t_{ij}$ as
\begin{eqnarray}
t_{ij}&=&  \nonumber
t_{ij}^{AA}  ,~{\rm if} \quad i\in A,\quad j\in A \\ \nonumber
& &t_{ij}^{BB}  ,~{\rm if} \quad i\in B,\quad j\in B \\ \nonumber
& &t_{ij}^{AB}  ,~{\rm if} \quad i\in A,\quad j\in B \\ 
& &t_{ij}^{BA}  ,~{\rm if} \quad i\in B,\quad j\in A,
\label{eq:}
\end{eqnarray}
where $t_{ij}$ depends on the type of ion occupying sites $i$ and $j$.
For off-diagonal disorder BEB~\cite{PhysRevB.4.2412} showed the scalar CPA equation
becomes a $2\times2$ matrix equation, with corresponding
AA, AB, BA, and BB matrix elements. 
In momentum space, if there is only near-neighbor hopping between all ions, the
bare dispersion can be written as (the under-bar denotes matrices)
\begin{eqnarray}
\underline{\varepsilon_k} & = \left(\begin{array}{cc} 
t^{AA} & t^{AB} \\ [1.0em]
t^{BA} & t^{BB} \end{array}\right) \varepsilon_k
\label{Eq: dispersion}
\end{eqnarray}
where in three dimensions $\varepsilon_k=-2t(\cos(k_x) +  \cos(k_y) +  \cos(k_z))$ with $4t=1$ 
which sets our unit of energy, and $t^{AA}$, $t^{BB}$, $t^{AB}$, and $t^{BA}$ are unitless
prefactors.

The BEB approach is local by construction,
hence all non-local disorder induced correlations are neglected.~\cite{PhysRevB.4.2412}
In order to take into account non-local physics, we extend the BEB formalism to a 
finite cluster using the DCA scheme. Here in the following, we present the algorithm and details of 
our non-local DCA extension of the BEB formalism for off-diagonal disorder. 
Just as in the DCA scheme,~\cite{m_jarrell_01a} the first Brillouin zone is divided 
into coarse-grained cells with centers $K$ surrounded by points $\tilde{k}$ within 
the cell so that an arbitrary $k=K+\tilde{k}$. 

For a  given DCA $K$-dependent effective medium 
hybridization $\underline{\Delta(K,\omega)}$ matrix 
\begin{equation}
\begin{split}
\underline{\Delta(K,\omega)}=
\left(\begin{array}{cc}
\Delta^{AA}(K,\omega) & \Delta^{AB}(K,\omega)\\
\Delta^{BA}(K,\omega) & \Delta^{BB}(K,\omega)
\end{array}\right)
\label{eq:-2-1}
\end{split}
\end{equation}
we solve the cluster problem, usually in real space. 
A set of stochastically generated random configuration of disorder potentials $v_i$ is used to 
calculate the disorder averaged cluster Green function
$ \overline{G}_{c}(K,\omega)$, 

\begin{equation}
\left(\underline{G_{c}}\right)_{ij}=\langle \left(\underline{\omega}-\underline{\overline{t'}}-
\underline{\Delta'}-\underbar{V}\right)_{ij}^{-1} \rangle
\label{eq:5}
\end{equation}
 where $\langle ... \rangle$ 
denotes disorder averaging and $\underbar{V}$ is a diagonal matrix for the disorder site potential. 
The primes stand for the configuration dependent Fourier transform (FT) components 
of the hybridization and hopping, respectively. I.e.,
\begin{subequations}
\begin{equation}
\underline{\Delta}'_{ij}  =\begin{cases}
FT(\Delta(K,\omega)^{AA})  ,~{\rm if} \quad i\in A,\quad j\in A\\
FT(\Delta(K,\omega)^{BB})  ,~{\rm if} \quad i\in B,\quad j\in B\\
FT(\Delta(K,\omega)^{AB})  ,~{\rm if} \quad i\in A,\quad j\in B\\
FT(\Delta(K,\omega)^{BA})  ,~{\rm if} \quad i\in B,\quad j\in A\end{cases}
\label{eq:-5}
\end{equation}
and 
\begin{equation}
\overline{\underline{t}}'_{ij}=\begin{cases}
FT(\overline{\epsilon}(K)^{AA})  ,~{\rm if} \quad i\in A,\quad j\in A\\
FT(\overline{\epsilon}(K)^{BB})  ,~{\rm if} \quad i\in B,\quad j\in B\\
FT(\overline{\epsilon}(K)^{AB})  ,~{\rm if} \quad i\in A,\quad j\in B\\
FT(\overline{\epsilon}(K)^{BA})  ,~{\rm if} \quad i\in B,\quad j\in A\end{cases}
\label{eq:-6}
\end{equation}
with 
\begin{eqnarray}
\underline{\overline{\epsilon}(K)} & = \left(\begin{array}{cc} 
t^{AA} & t^{AB} \\ [1.0em]
t^{BA} & t^{BB} \end{array}\right) \frac{N_{c}}{N}\sum_{\tilde{k}} \varepsilon_{k},
\label{Eq: eps_bar}
\end{eqnarray}
\end{subequations}
where $\underline{\Delta}'_{ij}$ and $\overline{\underline{t}}'_{ij}$
are $N_{c}\times N_{c}$ real-space matrices (where $N_c$ is the cluster size),
and e.g., $FT(\Delta(K,\omega)^{AA})=\sum_K \Delta(K,\omega)^{AA} e^{iK(r_i-r_j)}$.
The hopping can be long ranged, but since they
are coarse-grained quantities are effectively limited
to the cluster. Physically, $\underline{\Delta}'_{ij}$ represents
the hybridization between sites $i$ and $j$ which
is configuration dependent. For example, the AA component of the hybridization
corresponds to both A species occupying site $i$ and $j$, while the AB component
means  that site $i$ is occupied by an A atom and site $j$ by a B atom. 
The interpretation of the hopping matrix is the same as for the hybridization function.

In the next step, we form the $2N_c \times 2N_c$ disorder averaged Green function 
\begin{eqnarray}
\left\langle G_{c}(\omega)_{ij} \right\rangle &= \left(\begin{array}{cc}
\left\langle G_{c}^{AA}(\omega) \right\rangle_{ij} \quad \left\langle G_{c}^{AB}(\omega) \right\rangle_{ij}\\ [1.0em]
\left\langle G_{c}^{BA}(\omega) \right\rangle_{ij} \quad \left\langle G_{c}^{BB}(\omega) \right\rangle_{ij}
\end{array}\right).
\label{eq:10}
\end{eqnarray}
This may be done by assigning the components according to the occupancy of the sites $i$ and $j$ 
\begin{eqnarray}
(G_{c}^{AA})_{ij} &=&(G_{c})_{ij}\: ~{\rm if} \quad i\in A,\quad j\in A\nonumber \\
(G_{c}^{BB})_{ij} &=&(G_{c})_{ij}\: ~{\rm if} \quad i\in B,\quad j\in B\nonumber \\
(G_{c}^{AB})_{ij} &=&(G_{c})_{ij}\: ~{\rm if} \quad i\in A,\quad j\in B\nonumber \\
(G_{c}^{BA})_{ij} &=&(G_{c})_{ij}\: ~{\rm if} \quad i\in B,\quad j\in A\,
\label{eq:11}
\end{eqnarray}
with the other components being zero. Because only one of the four
matrix elements is finite for each disorder configuration (each site can be occupied by 
either $A$ or $B$ atom), only the sum of the
elements in Eq.~\ref{eq:10} is normalized as a conventional Green function.

Having formed the configuration dependent average Green function, we then Fourier transform 
to $K$-space (which also imposes translational symmetry) and obtain 
the K-dependent disorder averaged cluster Green function for each component of the matrix 

\begin{equation}
 G_{c}(K,\omega)=\left(\begin{array}{cc}
 G_{c}^{AA}(K,\omega)  & G_{c}^{AB}(K,\omega) \\ [1.0em]
 G_{c}^{BA}(K,\omega)  &  G_{c}^{BB}(K,\omega)\end{array}\right)\,.
\label{eq:-11}
\end{equation}

Once the cluster problem is solved, we calculate the coarse-grained lattice Green function as
\begin{eqnarray}
\bar{G}(K,\omega) & = & \left(\begin{array}{cc}   
\overline{G}^{AA}(K,\omega) & \overline{G}^{AB}(K,\omega) \nonumber \\ 
\overline{G}^{BA}(K,\omega) & \overline{G}^{BB}(K,\omega)\end{array}\right) \nonumber\\
& = & \frac{N_{c}}{N}\sum_{\tilde{k}}\Big(\underline{ G_{c}(K,\omega)}^{-1}+\underline{\Delta_{}(K,\omega)} \nonumber  \\
& - & \underline{\varepsilon_{k}}+\underline{\overline{\epsilon}(K+\tilde{k})}\Big)^{-1}.
\label{coarsegraining}
\end{eqnarray}
It is important to note that each component of the Green function matrix above does 
not have the normalization of a conventional, i.e., scalar, Green function.  Only the sum of the 
matrix components has the conventional normalization, so that $\overline{G}(K,\omega)\sim1/\omega$, with the 
total coarse grained lattice Green function being obtained as

\begin{eqnarray}
\overline{G}(K,\omega)&=& \overline{G}^{AA}(K,\omega)+\overline{G}^{BB}(K,\omega) \nonumber \\
                &+& \overline{G}^{AB}(K,\omega)+\overline{G}^{BA}(K,\omega).
\label{eq:-13}
\end{eqnarray}

Next, to construct the new DCA effective medium $\underline{\Delta(K,\omega)}$, 
we impose the BEB DCA $(2 \times 2)$ matrix self-consistency condition, requiring the
disorder averaged cluster 
and the coarse-grained  lattice Green functions to be equal
\begin{equation}
\underline{G_c(K,\omega)}=\underline{\bar{G}(K,\omega)} \,.
\end{equation}
This is equivalent to a system of three coupled scalar equations
\begin{subequations}
\begin{eqnarray}
 \overline{G}^{AA}(K,\omega) &=& G_{c}^{AA}(K,\omega), \\
\overline{G}^{BB}(K,\omega) &=& G_{c}^{BB}(K,\omega), \quad \textnormal{and} \\
 \overline{G}^{AB}(K,\omega) &=& G_{c}^{AB}(K,\omega).
\end{eqnarray}
\end{subequations}
Note $\overline{G}^{BA}(K,\omega)=\overline{G}^{AB}(K,\omega)$ automatically.

We then close our self-consistency loop by updating the corresponding hybridization functions 
for each components as
\begin{eqnarray}
\Delta_{n}^{AA}(K,\omega) &=& \Delta_{o}^{AA}(K,\omega) \nonumber \\
                            &+& \xi\left(G_{c}^{-1}(K,\omega)^{AA}-\overline{G}^{-1}(K,\omega)^{AA}\right) \nonumber \\ 
\Delta_{n}^{BB}(K,\omega) &=& \Delta_{o}^{BB}(K,\omega)  \nonumber \\
                            &+& \xi\left( G_{c}^{-1}(K,\omega)^{BB}-\overline{G}^{-1}(K,\omega)^{BB}\right) \nonumber \\ 
\Delta_{n}^{AB}(K,\omega) &=& \Delta_{o}^{AB}(K,\omega) \nonumber \\
                            &+& \xi\left( G_{c}^{-1}(K,\omega)^{AB}-\overline{G}^{-1}(K,\omega)^{AB}\right) \nonumber \\
\Delta_{n}^{BA}(K,\omega) &=& \Delta_{n}^{AB}(K,\omega)\label{eq:4}
\end{eqnarray}
where `o' and `n' denote old and new respectively, and $\xi$ is a linear mixing parameter $0<\xi< 1$.
We then iterate the above steps until convergence is reached.

There are two limiting cases of the above formalism which we carefully checked numerically.
In the limit of $N_{c}=1$, we should recover the original BEB result. Here the 
cluster Green function loses its $K$ dependence, so that
\begin{equation}
\begin{split}
\left(\begin{array}{cc}
G_{c}^{AA}(\omega) & 0\\  
0 & G_{c}^{BB}(\omega)
\end{array}\right)=  \\
\frac{1}{N}\sum_{{k}}\left( \underline{G_{c}(\omega)}^{-1}+
\underline{\Delta(\omega)}-\underline{\varepsilon(k)}\right)^{-1}
\label{eq:-14}
\end{split}
\end{equation}
which is the BEB self-consistency condition. 
Here we used that $\overline{\epsilon}(K)=0$ for N$_c=1$.
The second limiting case is when there is only diagonal disorder so that
$t^{AA}=t^{BB}=t^{AB}=1$. In this case the above formalism reduces to
the original DCA scheme. We have  verified numerically both these limits.

\subsection{Typical medium theory with off-diagonal disorder}
\label{sec:TMDCA}

To address the issue of electron localization, we recently developed the 
typical medium dynamical cluster approximation (TMDCA) and applied it to the three-dimensional Anderson 
model.~\cite{PhysRevB.89.081107} In Ref.~\onlinecite{PhysRevB.89.081107} we have confirmed 
that the typical density of states vanishes for states which are localized and it is finite for 
extended states.  In the following we generalize our TMDCA analysis to systems with off-diagonal disorder
to address the question of localization and the mobility edge in such models.

First, we would like to emphasize that the crucial difference between TMDCA~\cite{PhysRevB.89.081107} 
and the standard DCA~\cite{m_jarrell_01a} procedure is the way the disorder averaged cluster Green 
function is calculated. In the TMDCA analysis instead of using the algebraically averaged cluster 
Green function in the self-consistency loop, we calculate the typical (geometrically) averaged 
cluster density of states
\begin{equation}
\rho^c_{typ}(K,\omega)=e^{\frac{1}{N_c}\sum_{i}\langle\ln\rho_{ii}(\omega)\rangle}\left\langle \frac{-\frac{1}{\pi}
\Im G_{c}(K,\omega)}{\frac{1}{N_c}\sum_{i}(-\frac{1}{\pi}\Im G_{ii}(\omega))}\right\rangle,
\label{eq:-17}
\end{equation}
with the geometric averaging being performed over the local density of states 
$\rho^c_{ii}(K,\omega)=-\frac{1}{\pi}\Im G_{ii}(w)$ only.  Using this $\rho^c_{typ}(K,\omega)$ the 
cluster averaged typical Green function is constructed via a Hilbert transform
\begin{equation}
 G_{c}(K,\omega) =\int d\omega'\frac{\rho^c_{typ}(K,\omega')}{\omega-\omega'}.
\label{eq:-18-1}
\end{equation}

In the presence of off-diagonal disorder, following BEB, the typical density of states becomes a 
$2 \times 2$ matrix, which we define as
\begin{widetext}
\begin{eqnarray} 
\addtolength{\jot}{1em}
\underline{\rho^c_{typ}(K,\omega)} & = \exp\left(\dfrac{1}{N_c} \sum_{i=1}^{N_c} \left\langle \ln \rho_{ii} (\omega) \right\rangle\right) 
\times & \left(\begin{array}{cc}
\left\langle \dfrac{-\dfrac{1}{\pi}\Im G_{c}^{AA}(K,\omega)}{\frac{1}{N_c} \sum_{i=1}^{N_c}(-\dfrac{1}{\pi}\Im G_{ii}(\omega))}\right\rangle  & \left\langle \dfrac{-\frac{1}{\pi}\Im G_{c}^{AB}(K,\omega)}{\frac{1}{N_c} \sum_{i=1}^{N_c}(-\frac{1}{\pi}\Im G_{ii}(\omega))}\right\rangle \\ [1.8em]
\left\langle \dfrac{-\dfrac{1}{\pi}\Im G_{c}^{BA}(K,w)}{\frac{1}{N_c} \sum_{i=1}^{N_c}(-\dfrac{1}{\pi}\Im G_{ii}(\omega))}\right\rangle  & \left\langle \dfrac{-\frac{1}{\pi}\Im G_{c}^{BB}(K,\omega)}{\frac{1}{N_c} \sum_{i=1}^{N_c}(-\frac{1}{\pi}\Im G_{ii}(\omega))}\right\rangle \end{array}\right).
\label{rhotyp_BEB}
\end{eqnarray}
\end{widetext}
Here the scalar prefactor depicts the local typical (geometrically averaged) density of states, while
the matrix elements are linearly averaged over the disorder. Also notice that 
the cluster Green function $(\underline{G_c})_{ij}$ and its components $G_c^{AA}$, $G_c^{BB}$ 
and $G_c^{AB}$ are defined in the same way as in Eqs. (\ref{eq:5}-\ref{eq:11}).

In the next step, we construct the cluster average Green function $G_c(K,\omega)$ by performing Hilbert transform 
for each component
\begin{eqnarray} 
\addtolength{\jot}{1em}
\underline{ G_c(K,\omega)} & =  & \left(\begin{array}{cc}
\int d\omega'\frac{\rho_{typ}^{AA}(K,\omega')}{\omega-\omega'}  
& \int d\omega'\frac{\rho_{typ}^{AB}(K,\omega')}{\omega-\omega'} \\ [1.8em]
\int d\omega'\frac{\rho_{typ}^{BA}(K,\omega')}{\omega-\omega'}
& \int d\omega'\frac{\rho_{typ}^{BB}(K,\omega')}{\omega-\omega'}
\end{array}\right).
\label{Gtyp_BEB}
\end{eqnarray}

Once the disorder averaged cluster Green function $G_c(K,\omega)$  is obtained from
Eq.~\ref{Gtyp_BEB}, the self-consistency steps are the same as in the procedure for the off-diagonal disorder
DCA described in the previous section: we calculate the coarse-grained lattice Green function
using Eq.~\ref{coarsegraining} which is then used
to update the hybridization function with the effective medium via Eq.~\ref{eq:4}.
 
The above set of equations provide us with the generalization of the TMDCA 
scheme for both diagonal and off-diagonal disorder which we test numerically in the following sections.
Also notice that for $N_c=1$ with only diagonal disorder ($t^{AA}=t^{BB}=t^{AB}=t^{BA}$)  the above procedure 
reduces to the local TMT scheme. In this case, the diagonal elements of the matrix
in Eq.~\ref{rhotyp_BEB} will contribute $c_A$ and $c_B$, respectively, with the off-diagonal elements
being zero (at N$_c=1$ the off-diagonal terms vanish because
at a given site the can be either $A$ or $B$ atom only). Hence, the typical density reduces to the local scalar prefactor only, which
has exactly the same form as in the local TMT scheme.

%%%%%%%%%%%%%%%%%%%%%%%%%%%%%%%%%%%%%%%%%%%%%%%%%%%%%%%%%%%%%%%%%%%%%%%%%%%%%%%%%%%%%%%
\section{Results and Discussion}
\label{sec:Results_Disc}
To illustrate the generalized DCA and TMDCA algorithms described above, 
we present our results for the effects of diagonal 
and off-diagonal disorder in a generalized Anderson Hamiltonian (Eq.~\ref{eq:1}) for a three 
dimensional system with binary disorder distribution ($V_A = -V_B$) and random hopping 
($t^{AA} \neq t^{BB}$, $t^{AB} = t^{BA}$) with other parameters as specified. The results are 
presented and discussed in Subsections~\ref{sec:Results_DCA} and \ref{sec:Results_TMDCA}.  

\subsection{DCA results for diagonal and off-diagonal disorder}
\label{sec:Results_DCA}

\begin{figure}[t!]
 \includegraphics[trim = 0mm 0mm 0mm 0mm,width=1\columnwidth,clip=true]{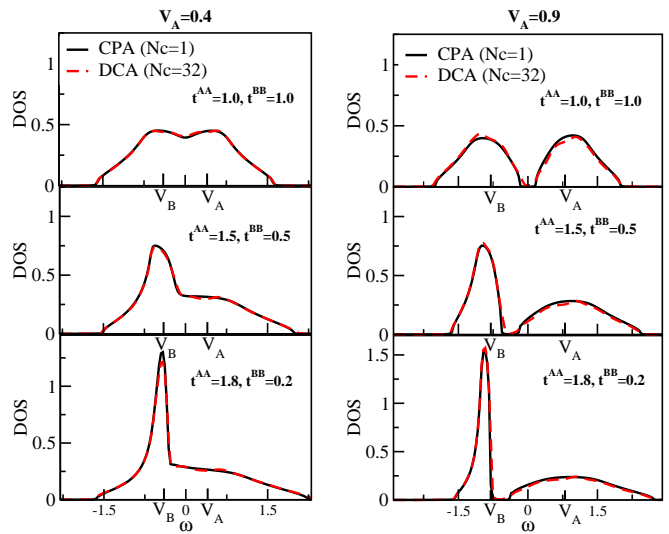}
\caption{(Color online). The effect of off-diagonal disorder on the average density of 
states calculated in the DCA scheme with $N_c=32$. Our DCA results for $N_c=1$ corresponds 
to a single site CPA BEB scheme. We consider two values of local disorder potential below ($V_A=0.4$) and 
above ($V_A=0.9$) the band-split limit, and examine the effect of changing the off-diagonal hopping strength (which
amounts to a change in the non-local potential).  
We start with the diagonal disorder case $t^{AA}=t^{BB}=t^{AB}=1.0$ and then 
consider two off-diagonal disorder cases: $t^{AA}=1.5, t^{BB}=0.5$
and $t^{AA}=1.8, t^{BB}=0.2$, respectively. We fix $t^{AB}=t^{BA}=0.5(t^{AA}+t^{BB})$ and $c_A=0.5$. 
For this parameter range of off-diagonal disorder, we do 
not observe a significant difference between the 
CPA ($N_c=1$) and the DCA ($N_c=32$) results indicating that non-local inter-site correlations are weak.}
\label{fig:Fig1} 
\end{figure}

The effect of off-diagonal disorder on the average density of states (DOS) calculated within the 
DCA ($N_c=32$) is presented in Fig.~\ref{fig:Fig1}. The DOS we present in our results is a 
local density of states calculated as 
\begin{eqnarray}
DOS(\omega) & = & -\frac{1}{\pi N_c}\sum_{K=1}^{N_c} 
\Big(\Im \overline{G}^{AA}(K,\omega)+\Im \overline{G}^{AB}(K,\omega) \nonumber  \\
& + & \Im \overline{G}^{BA}(K,\omega)+\Im \overline{G}^{BB}(K,\omega)\Big).
\end{eqnarray}
Notice that our TMDCA procedure for $N_c=1$ reduce to 
the original CPA-like BEB.  For a fixed concentration $c_A=0.5$, we examine the 
effects of off-diagonal disorder at two fixed values of the diagonal disorder potential $V_{A}=0.4$ 
(below the split-band limit) and $V_{A}=0.9$ (above the split-band limit).
The off-diagonal randomness is modeled by changes in the hopping amplitudes $t^{AA},t^{BB}$
with $t^{AB}=0.5(t^{AA}+t^{BB})$.
For a diagonal disorder case (top panel of Fig.~\ref{fig:Fig1}) with $t^{AA} = t^{BB} = t^{AB} = t^{BA}$
we have two subbands contributing equally to the total DOS.  
While as shown in the middle and bottom panels, the change in the strength of the
off-diagonal disorder leads to dramatic changes in the DOS. 
An increase of the AA hoping results in the broadening 
of the AA subband with the development of a resonance peak at the BB subband.
For this parameter range both the DCA ($N_c=32$ ) and CPA ($N_c=1$) provide about the same results indicating that 
disorder-induced non-local correlations are negligible.

\begin{figure}[tbh]
\includegraphics[trim = 0mm 0mm 0mm 0mm,width=1\columnwidth,clip=true]{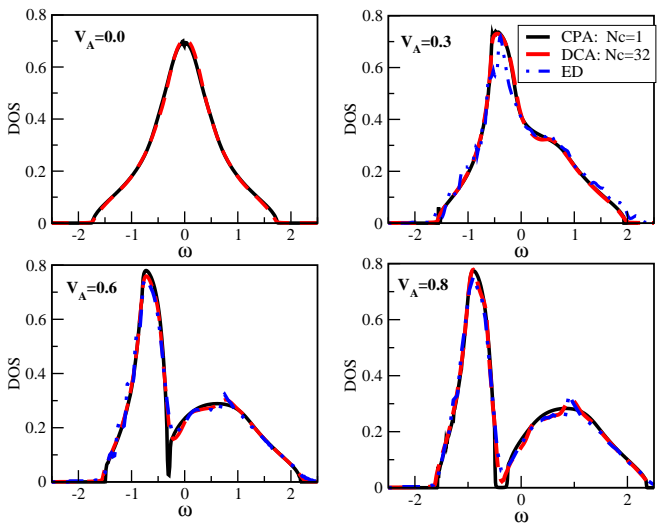}
\caption{(Color online). The effect of an increasing diagonal disorder potential $V_A$ for a fixed off-diagonal 
disorder with $t^{AA}=1.5$, $t^{BB}=0.5$, $t^{AB}=0.5 (t^{AA}+t^{BB}),$ $c_{A}=0.5.$ on the 
the average density of states calculated with our modified DCA scheme. Results are obtained for $N_c=1$ 
(corresponding to the CPA) and $N_c=32$ cluster sizes.
We also compare our DCA average DOS with the DOS obtained using exact diagonalization (ED)
for a $12 \times 12 \times 12$ cubic lattice cluster.
For ED results, we used $\eta=0.01$ broadening in frequency.
\label{fig:Fig2}}
\end{figure}

In Fig.~\ref{fig:Fig2}, we show the average density of states calculated for
fixed off-diagonal-disorder parameters and different diagonal disorder potentials $V_{A}$.
We again compare local CPA ($N_c=1$) and the DCA ($N_c=32$) results.
To benchmark our off-diagonal extension of the DCA,  we compare our results with 
those obtained from exact diagonalization. For small $V_{A}$, there is no difference between the CPA 
($N_c=1$) and the DCA ($N_c=32$) results. As local potential $V_{A}$ is increased,  noticeable 
differences start to develop. We can see that for larger $V_{A}$ a gap starts to open and is more
dramatic in the CPA scheme. While in the DCA ($N_c=32$) this gap is partially filled due to the incorporation of non-local 
inter-site correlations which are missing in the CPA. Furthermore, the DOS obtained from the       
DCA procedure provides finer structures which are in basic agreement with the DOS calculated
with exact diagonalization for a cluster of size $12 \times 12 \times 12$.   
The agreement we get with ED results is a good indication of the the accuracy of our extension of the
 DCA to off-diagonal disorder.

\subsection{Typical medium finite cluster analysis of diagonal and off-diagonal disorder}
\label{sec:Results_TMDCA}

\begin{figure}[t!]
 \centering{} \includegraphics[trim = 0mm 0mm 0mm 0mm,width=1\columnwidth,clip=true]{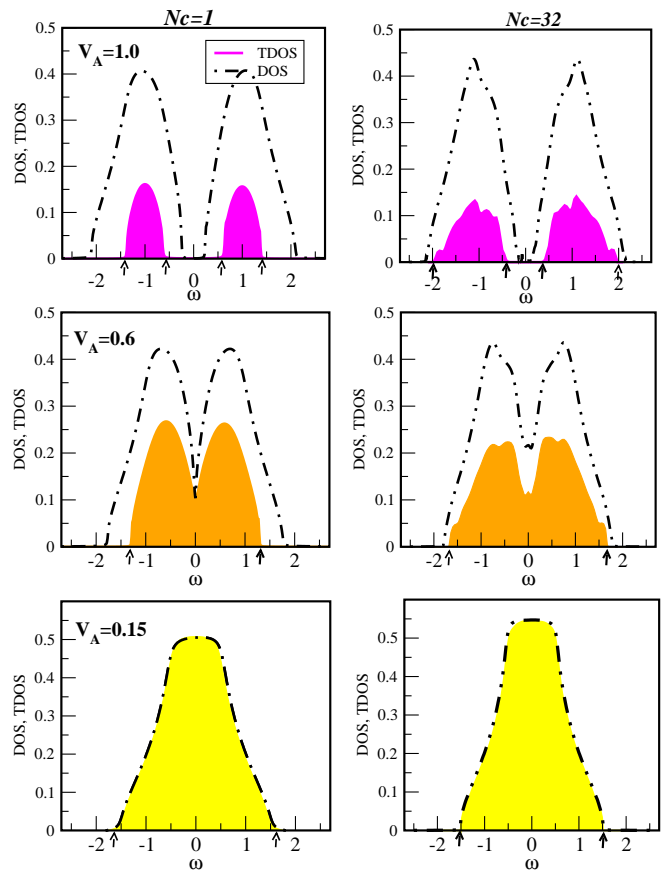}
 \caption{(Color online). Diagonal disorder case: 
the average density of states (dash-dotted line) calculated within the DCA and the typical density
of states (shaded regions) calculated within the TMDCA for diagonal disorder $t^{AA}= t^{BB}= 
t^{AB}=t^{BA}=1, c_A=0.5$ and various values of the local potential $V_A=-V_B$ for cluster sizes 
$N_c=1$ and $N_c=32$.  The shaded regions represent the TDOS which is finite for the extended 
states and zero when the states are localized.  The mobility edges extracted 
from the vanishing of the TDOS are marked by the arrows. The extended states region with a finite TDOS 
is always narrower for $N_c=1$ as compared to the results of a $N_c=32$ cluster, indicating that a single 
site TMT tends to overemphasize the localized states.
\label{fig:Fig3}}
\end{figure}

To characterize the Anderson localization transition, we now explore 
the typical density of states (TDOS) calculated within our extension of the TMDCA
presented in Sec.~\ref{sec:TMDCA}. In the typical medium analysis, the TDOS serves as the order parameter 
for the Anderson localization transition.
In particular, the TDOS is finite for extended states and zero for states which are localized.

First we consider the behavior of the  TDOS and compare it with the average DOS for diagonal disorder.
In Fig.~\ref{fig:Fig3} we show our results for $N_c=1$ (left panel)
and $N_c=32$ (right panel). Notice that $N_c=1$ results for TDOS correspond to the single-site TMT of 
Dobrosavljevi\'{c} \etal,~\cite{Vlad2003} and for average DOS they correspond to the ordinary CPA.
As expected,~\cite{Vlad2003, PhysRevB.89.081107} for small disorder ($V_A=0.15$) there is not much difference between the DCA ($N_c=32$)
and the TMDCA ($N_c=32$) or between the CPA and TMT for $N_c=1$ results. However, there are subtle differences between 
the results for finite $N_c=32$ and single site $N_c=1$ clusters due to incorporation of 
spatial correlations. As the disorder strength $V_A$ is increased ($V_A=0.6$), the typical density of states
(TDOS) becomes smaller than the average DOS and is broader for the larger cluster.
Moreover, the finite cluster  introduce  features in the DOS which are missing in the local 
$N_c=1$ data. Regions where the TDOS is zero while the average DOS is finite 
indicate Anderson localized states, separated by the mobility edge (marked by arrows).
For $N_c>1$ these localized regions are wider which indicates that the localization edge 
is driven to higher frequencies. This is a consequence of the tendency of non-local corrections to suppress 
localization.  For even larger disorder $V_A=1$, a gap opens in both the TDOS and the average DOS leading to the formation 
of four localization edges, but again the region of extended states is larger 
for the finite cluster, indicating that local TMT ($N_c=1$) tends to underestimate the extended states
region.
\begin{figure}[tbh]
\includegraphics[trim = 0mm 0mm 0mm 0mm,width=1\columnwidth,clip=true]{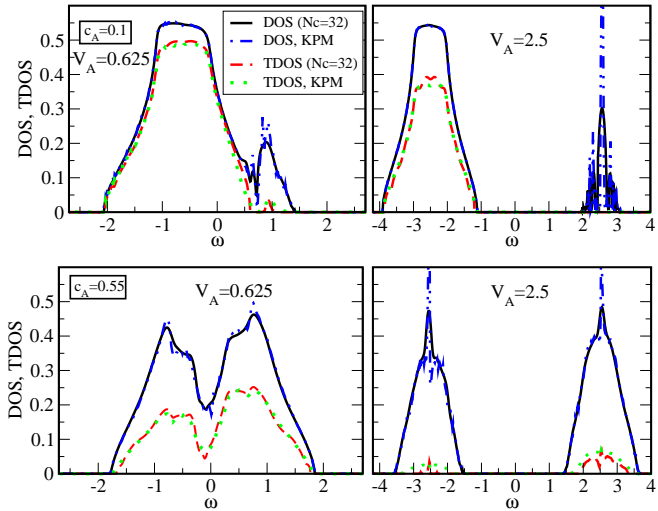}
 \caption{(Color online). Diagonal disorder case: Comparison of the average and typical DOS calculated with 
the DCA/TMDCA and Kernel polynomial methods (KPM)~\cite{Schubert} for diagonal disorder 
with $t^{AA}= t^{BB}= t^{AB}=t^{BA}=1$ at various values of local potential $V_A$ and 
concentrations $c_A$ for cluster size $N_c=32$.   The kernel polynomial method used
$2048$ moments on a $48^3$ cubic 
lattice, and $200$ independent realizations generated with $32$ sites randomly sampled from each realization.}
\label{fig:TDOS_KPM_comparison}
\end{figure}

To further benchmark our results for the diagonal disorder, we show in Fig.~\ref{fig:TDOS_KPM_comparison}
a comparison of the average and typical DOS calculated with the DCA and the TMDCA ($N_c=32$) 
as compared with the kernel polynomial method (KPM).~\cite{KPM_review_2006, Schubert}
In the KPM analysis, instead of diagonalizing the Hamiltonian directly, the local DOS is expressed in term of
an infinite series of Chebyshev polynomials. In practice, the truncated series leads to
Gibbs oscillations. The KPM damps these oscillations by a modification
of the expansion coefficients. Following previous studies on the Anderson model, the 
Jackson kernel is used.~\cite{KPM_review_2006}  As it is evident from the plots, our 
TMDCA results reproduced those from the KPM nicely showing that our formalism 
offers a systematic way of studying the Anderson localization transition in binary 
alloy systems. Such good agreement indicates a successful benchmarking of the TMDCA 
method.~\cite{PhysRevB.89.081107}

\begin{figure}[tbh]
\centering{} \includegraphics[trim = 0mm 0mm 0mm 0mm,width=1\columnwidth,clip=true]{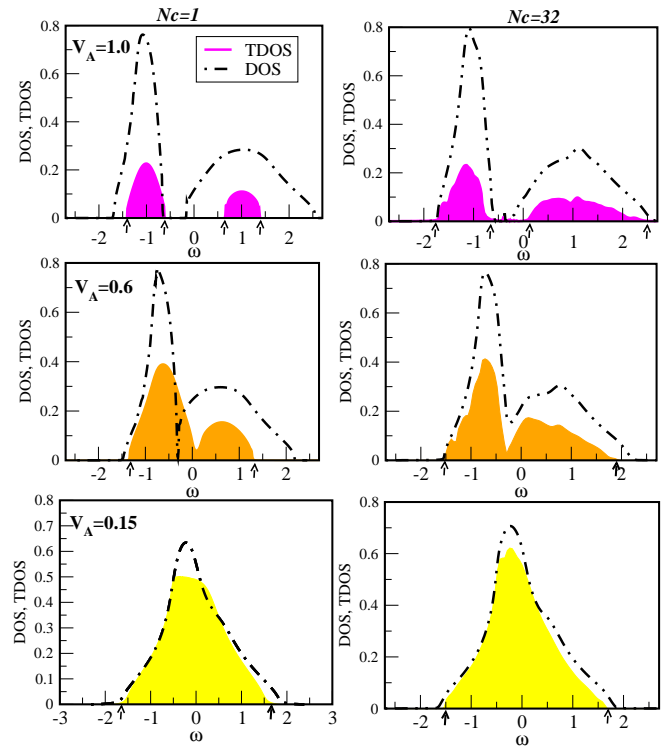}
\caption{(Color online). Off-diagonal disorder case: the evolution of the average density of states (dash-dotted line) and 
the typical density of states (shaded regions) for 
various values of the local potential $V_A$ with off-diagonal disorder parameters: $t^{AA}=1.5$,
$t^{BB}=0.5$, $t^{AB}=0.5(t^{AA}+t^{BB})$, and $c_{A}=0.5$. The left panel is for $N_c=1$ and 
the right panel for $N_c=32$. The mobility edges are extracted as described in Fig.~\ref{fig:Fig3}.
\label{fig:Fig4}}

\end{figure}

Next, we explore the effects of the off-diagonal disorder. In Fig.~\ref{fig:Fig4}, we compare 
the typical TDOS from the TMDCA and average DOS from the DCA for several values of the diagonal disorder 
strength $V_A$ at fixed off-diagonal disorder amplitudes $t^{AA}=1.5$, $t^{BB}=0.5$, $t^{AB}=1.0$.
To show the effect of a finite cluster with non-local correlations, we present data for single site $N_{c}=1$ and finite 
cluster $N_{c}=32$. The TMT ($N_c=1$) again underestimates the extended states 
regime by having a narrower TDOS as compared to $N_c=32$. For small disorder $V_{A}$ both the DOS 
and the TDOS are practically the same. However, as $V_{A}$ increases, 
significant differences start to emerge. Increasing $V_A$ leads to the gradual opening of the gap
which is more pronounced in the $N_c=1$ case and for smaller disorder $V_{A}=0.6$  is partially filled 
for the $N_c=32$ cluster.  As compared to the diagonal disorder case of  Fig.~\ref{fig:Fig3},
the average DOS and TDOS become
asymmetric with respect to zero frequency due to the off-diagonal randomness.
\begin{figure}[t!]
\includegraphics[trim = 0mm 0mm 0mm 0mm,width=1\columnwidth,clip=true]{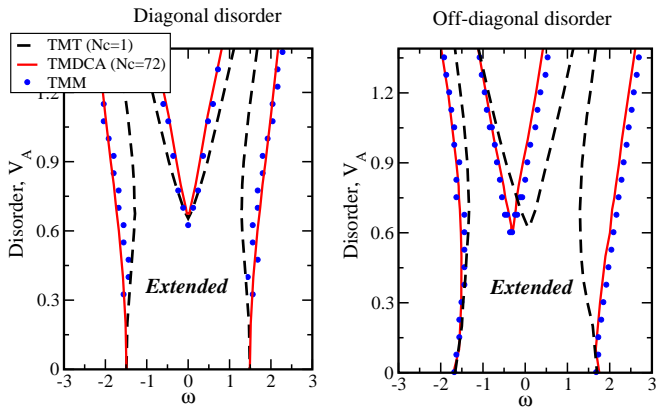}
\caption{(Color online). Disorder-energy phase diagram for the diagonal disorder case (left panel) and 
for the off-diagonal disorder case (right 
panel).  Parameters used for the diagonal disorder case are: $t^{AA}=t^{BB}=t^{AB}=1.0$, and $c_A=0.5$.
For the off-diagonal disorder case: $t^{AA}=1.5$, $t^{BB}=0.5$, $t^{AB}=1.0$, and $c_A=0.5$.  We compare the 
mobility edges obtained from the TMT $N_c=1$ (dash line), TMDCA $N_c=72$ 
(solid line), and the transfer-matrix method (TMM) (dotted line).
The single site $N_c=1$ strongly underestimates the extended states region in both the
diagonal and especially in the off-diagonal disorder. The mobility edges obtained from the finite cluster 
TMDCA ($N_c=72$) show good agreement with those obtained from the TMM, in contrast to single 
site TMT. See text for parameters and details of the TMM implementation.
\label{fig:Fig5}}
\end{figure}

In Fig.~\ref{fig:Fig5} we present the disorder-energy phase diagram for both diagonal (left panel)
and off-diagonal (right panel) disorder calculated using the single TMT ($N_c=1$) and the non-local TMDCA ($N_c=72$).
To check the accuracy of the mobility edge trajectories extracted from our typical medium analysis, 
we compare our data with the results obtained with the transfer matrix method (TMM).

The TMM~\cite{MacKinnonKramer1983,Kramer2010,Markos-review-2006} is 
a well established numerical method  for calculating the correlation length 
and determining the mobility edge of the disorder Anderson model. 
Its main advantage is in its capability of capturing the effects from
rather large system sizes. Thus, it provides good data for a finite size scaling analysis
to capture the critical points and the corresponding exponents. In our calculations,
the transmission of states down a three-dimensional bar of widths $M = [6,12]$ and length $L = 2
\times 10^4 M$ are studied by adding the products of the transfer matrices with random initial states.
The multiplication of transfer matrices is numerically unstable. To avoid 
this instability, we orthogonalized the transfer matrix product every five multiplications 
using a Lapack QR decomposition.~\cite{Slevin2014} The localization edge is obtained by calculating the 
Kramer-MacKinnon scaling parameter $\Lambda_M$.~\cite{MacKinnonKramer1983} This is a dimensionless
quantity which should be invariant at the critical point, that is,
$\Lambda_M$ scales as a constant for $M \rightarrow \infty$.~\cite{Kramer2010}
Thus, we determine the boundary of the localization transition vis-\`{a}-vis the critical 
disorder strength~\cite{plyushchay2003} by performing 
%locating the disorder strength where $\Lambda_M$ decreases with increasing $M$ on the 
%plot of $\Lambda_M$ versus $M$ to categorize the localized states, for example.  Here, this is determined by 
a linear fit to $\Lambda_M$ v. $M$ data: localized states will have a negative slope
and visa versa for extended states. 
The transfer-matrix method finite size effects are larger for weak disorder where the states decay slowly 
with distance and so have large values of $\Lambda_M$ that carry a large variance in the data.
Notice that the CPA and the DCA do not suffer such finite size effect limitation 
for small disorder and are in fact exact in this limit.

The mobility edges shown in Fig.~\ref{fig:Fig5} were extracted from the
TDOS, with boundaries being defined by zero TDOS. 
As can be seen in Fig.~\ref{fig:Fig5}, while the single-site TMT does not change much under the 
effect of off-diagonal disorder, the TMDCA results are significantly modified. 
The bands for a larger cluster become highly asymmetric with significant 
widening of the A subband. The local $N_c=1$ boundaries are narrower than those obtained for 
$N_c=72$ indicating that the TMT strongly underestimates
the extended states regime in both diagonal and off-diagonal disorder. 
On the other hand, comparing  the mobility edge boundaries for $N_c=72$ with those obtained 
using TMM, we find very good agreement. This again confirms the validity of our 
generalized TMDCA.

\begin{figure}[tbh]
\centering{} \includegraphics[trim = 0mm 0mm 0mm 0mm,width=1\columnwidth,clip=true]{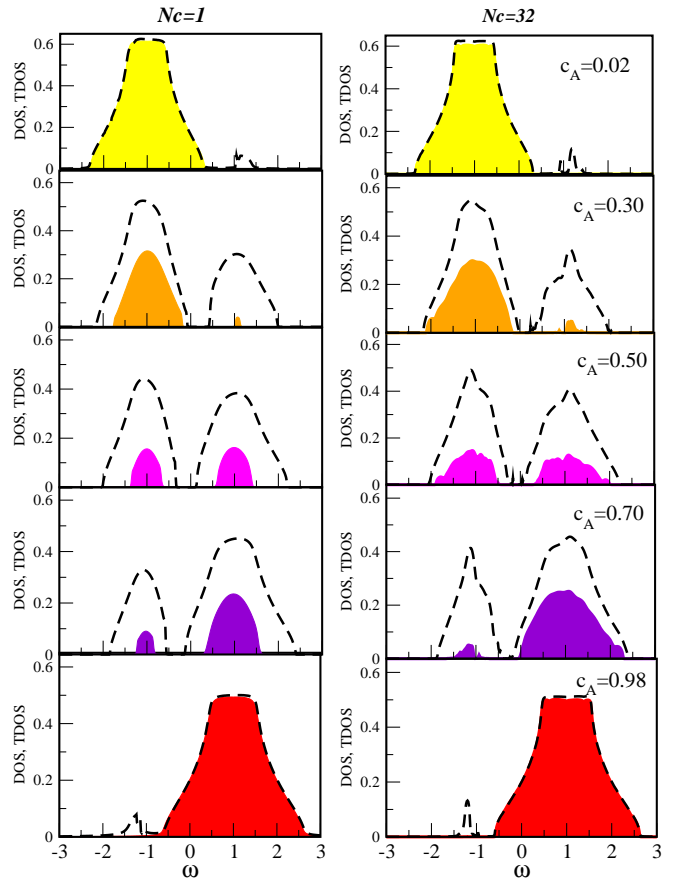}
\caption{(Color online). The average DOS (dot-dashed lines) and the typical DOS (shaded regions) 
for various values of the concentration $c_A$  with
off-diagonal disorder parameters $t^{AA}=1.1$, $t^{BB}=0.9$ and $t^{AB}=1.0$,
at fixed local potential $V_A=1.0$ for $N_c=1$ (left panel) and $N_c=32$ (right panel).}
%$c_A=0.02. 0.3, 0.5, 0.7, 0.98$ 
\label{fig:concentartion_tdos_ados}
\end{figure}

\begin{figure}[tbh]
\begin{raggedright}
\includegraphics[scale=0.225]{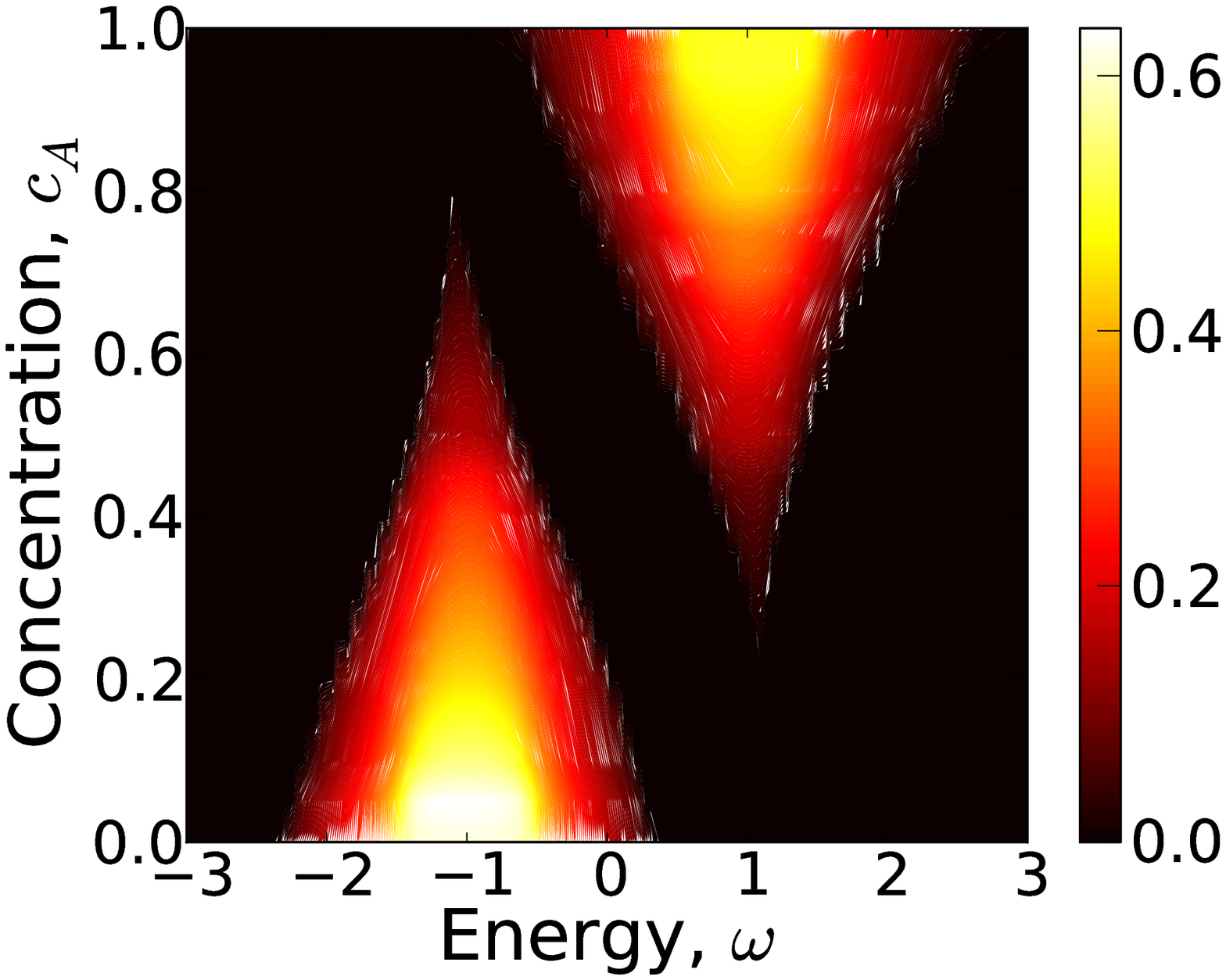} \includegraphics[scale=0.225]{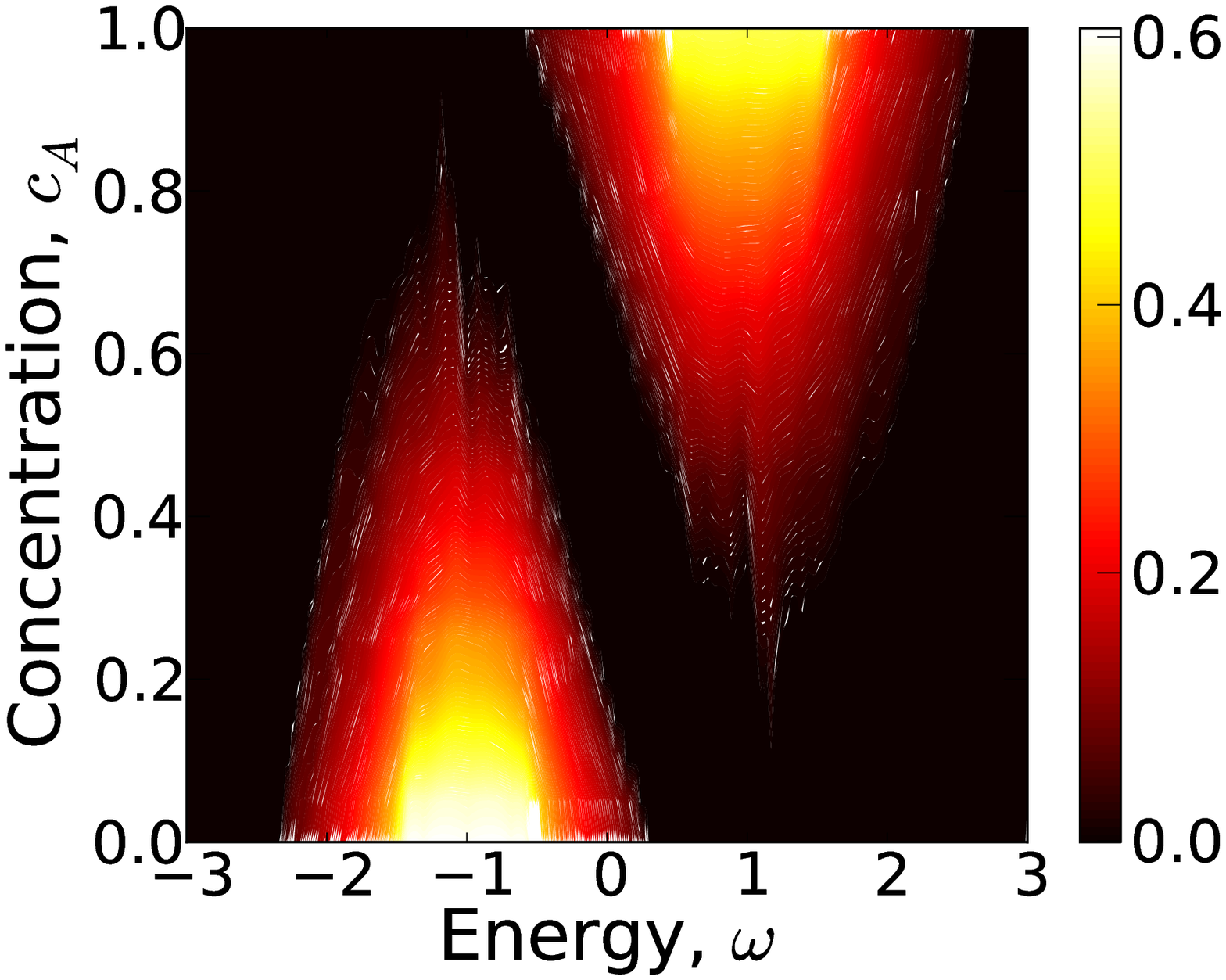} 
\par\end{raggedright}
\caption{(Color online). The evolution of the typical density of states for $N_c=1$ (left panel) and $N_c=32$ (right panel)
with the change in the concentration $0<c_A<1$ at fixed diagonal and off-diagonal disorder 
parameters: $t^{AA}=1.1$, $t^{BB}=0.9$, $t^{AB}=1.0$ and $V_A=1.0$ }
\label{fig:concentartion_contour}
\end{figure}

Next, we consider the effect of off-diagonal disorder at various concentrations $c_A$.  
In Fig.~\ref{fig:concentartion_tdos_ados},
we show the typical and average DOS for several values of $c_A$ calculated with the TMDCA and the DCA,
respectively. As expected, when $c_A\rightarrow 0$, we obtain a pure $B$ subband contribution (the top panel).
Upon gradual increase of the $c_A$ concentration, the number of states in the $A$ sub-band grows
until $B$-subband becomes a minority for $c_A>0.5$ and completely disappears at $c_A\rightarrow 1$ (the bottom panel).
Again, we see that a finite cluster $N_c=32$ provides a more accurate description 
(with finite details in DOS and broader regions of extended states in TDOS) in both average DOS and TDOS. 
The associated contour plots for the evolution of the TDOS in the concentrations range 
$0\leq c_A\leq1$ are shown in Fig.~\ref{fig:concentartion_contour}. The essence of these plots is 
to show the overall evolution of the typical and average DOS for a fixed local potential 
and off-diagonal disorder parameters as a function of 
the concentration $c_A$. In the limit of $c_A\rightarrow 0$, only the B-subband centered around $\omega=-V_A$ survives,
and for $c_A\rightarrow 1$, only the A-subband centered around $\omega=V_A$ is present. 
For intermediate concentrations, we clearly have contributions to the total typical 
density of states from both species, as expected. 
%Our results indicate that the discussed TMDCA scheme can be generalized to study the localization effects in multiband systems. 

Finally, we would like to comment on the possible further development 
of the presented scheme. After certain 
generalizations our current implementation of the typical medium dynamical cluster approximation 
for off-diagonal disorder 
can serve as the natural formalism for multiband (multiorbital)
systems.~\cite{Koepernik} Such an extension is crucial for studying disorder and localization effects in 
real  materials. Further development towards this direction
will be the subject of future publications.

%%%%%%%%%%%%%%%%%%%%%%%%%%%%%%%%%%%%%%%%%%%%%%%%%%%%%%%%%%%%%%%%%%%%%%%%%%%%%%%%%
\section{Conclusion}
\label{sec:conclusion}
A proper theoretical description of disordered materials requires the inclusion of both diagonal and off-diagonal randomness. 
In this paper,  we have extended the BEB 
single site CPA scheme to a finite cluster DCA that incorporates the effect of non-local 
disorder. Applying the generalized DCA scheme to a single band
tight binding Hamiltonian with configuration-dependent hopping amplitudes, we 
have considered the effects of non-local disorder and the interplay
of diagonal and off-diagonal disorder on the average density of states. 
By comparing our numerical results with those from exact diagonalization, we have
established the accuracy of our method.

To study the effect of disorder on electron localization and to determine the mobility 
edge in systems with both diagonal and off-diagonal randomness, we have also extended
our recently developed TMDCA to included off-diagonal randomness.
Within the TMDCA the typical DOS vanishes for localized states, and is finite 
for states which are extended. Employing the typical DOS as an order parameter for
Anderson localization, we have constructed the disorder-energy phase diagram for systems
with both diagonal and off-diagonal disorder. We have also demonstrated the inability of the single site
CPA and the TMT methods to capture accurately the localization and disorder effects in both
the average and the typical DOS, respectively.
Comparing our results with kernel polynomial, exact diagonalization, and transfer-matrix methods
we find  a remarkably good agreement with our extended DCA and TMDCA.
To the best of our knowledge, this is the first numerically accurate investigation of the Anderson localization
in systems with off-diagonal disorder within the framework of the typical medium analysis. 
We believe that the extended TMDCA scheme presents a powerful tool for 
treating both diagonal and off-diagonal disorder on equal footing, and can be easily extended to
study localization in multi-band systems.

\begin{acknowledgments}

We thank A. Gonis for useful discussion and directing us to the BEB formalism, we also 
thank Shuxiang Yang, Wei Ku, and Tom Berlijn for useful discussions.  This work is supported 
by DOE SciDAC grant DE-FC02-10ER25916 (MJ) and BES CMCSN grant DE-AC02-98CH10886 (HT).  
Additional support was provided by NSF EPSCoR Cooperative Agreement No. EPS-1003897 (HT, CE, 
CM), and NSF OISE-0952300 (JM). This work used the 
Extreme Science and Engineering Discovery Environment (XSEDE), which is supported by National 
Science Foundation grant number ACI-1053575,  the high performance computational resources provided by the 
Louisiana Optical Network Initiative (http://www.loni.org), and HPC@LSU computing.

\end{acknowledgments}

%\bibliography{paper_tmt-dca-odd}

\end{document}